\documentstyle[11pt,epsbox]{article}

\title{{\bf CP Violation in Neutrinoless Double Beta  Decay}}

\author{Takeshi FUKUYAMA \footnotemark[1],~~Kouichi MATSUDA  
\footnotemark[1
]~~ 
and Hiroyuki NISHIURA \footnotemark[2]  
                   \\
        $\ast$~~Department of Physics, \\
        Ritsumeikan University, Kusatsu, \\
        Shiga, 525 Japan \\
        $\dagger$~~ Department of General Education, \\
        Junior College of Osaka Institute of Technology, \\
        Asahi-ku, Osaka,535 Japan}

\date{}

\begin{document}

\maketitle



\begin{abstract}
We argue three-flavour neutrino mixing.  We consider the neutrinos as 
Majorana
particles and see how the neutrinoless double beta decay constrains the 
neutrino mixing
angles.  
Our formulation is widely valid and is applied  to the neutrino oscillation
experiment.
\\
PACS number(s): 14.60Gh 13.35.+s

\end{abstract}

\newpage



It is one of the most important problems in particle physics 
whether
the neutrinos 
have masses or not.  From the recent neutrino experiments 
\cite{Kamioka} 
\cite{Homestake} \cite{Gallex} \cite{Sage}, it becomes very probable 
that
the 
neutrinos have masses. However, if the neutrinos have masses, we must explain the 
reason why they are so tiny relative to the charged lepton masses.  Seesaw mechanism 
is one of the most promising candidates for such an
explanation. In this case, neutrinos become 
Majorana particles. 
\par
In this paper we consider the neutrinos as massive Majorana 
particles with three generations
and 
see how this point of view constrains physics of lepton sector.
\par
As is well known, the neutrino oscillation does not distinguish Majorana
neutrinos 
from Dirac ones. So, lets us first consider the neutrinoless double beta decay 
($(\beta\beta)_{0{\nu}}$) which occur only in the case of Majorana
neutrinos.  
\begin{center}
 Fig.1
\end{center}
The decay rate of $(\beta\beta)_{0{\nu}}$ is, in the absence of right-handed 
couplings, proportional to  the "averaged"  
mass defined by \cite{doi}
\begin{equation} 
\langle m_{\nu} \rangle\equiv |\sum _{j=1}^{3}U_{ej}^2m_j|.
\end{equation}
Here $U_{\alpha j}$ is the left-handed neutrino mixing 
matrix
which 
combines the weak eigenstate ($\alpha =e,\mu$ and $\tau$) to the mass 
eigenstate with mass \(m_j\) (j=1,2 and 3). 
It takes the following form in the case of Majorana 
neutrinos,
\begin{equation}
U=
\left(
\begin{array}{ccc}
c_1c_3&s_1c_3e^{i\beta}&s_3e^{i(\rho-\phi )}\\
(-s_1c_2-c_1s_2s_3e^{i\phi})e^{-i\beta}&c_1c_2-s_1s_2s_3e^{i\phi}&s_
2c_3e^{
i(\rho-\beta )}\\
(s_1s_2-c_1c_2s_3e^{i\phi})e^{-i\rho}&(-c_1s_2-s_1c_2s_3e^{i\phi})e^
{-i(\rho
 
-\beta )}&c_2c_3\\
\end{array}
\right).
\end{equation}
Here $c_j=cos\theta_j$, $s_j=sin\theta_j$ 
($\theta_1=\theta_{12},~\theta_2=\theta_{23},~\theta_3=\theta_{31}$),  
and beside $\phi$ ,appear the two additional CP violating phases $\beta$ and 
$\rho$ for Majorana particle.  
So $\langle m_{\nu} \rangle$ becomes 
\begin{equation}
\langle m_{\nu} \rangle=|m_1c_1^2c_3^2-m_2s_1^2c_3^2e^{-2i\beta 
'}-m_3s_3^2e^{-2i\rho '}|,
\end{equation}
where we have introduced 
\begin{equation}
\beta '\equiv {\pi\over 2}-\beta ,\quad \rho '\equiv {\pi\over
2}-(\rho-\phi ).
\end{equation}    
CP violation occurs in the presence of imaginary part in $\langle 
m_{\nu} \rangle$, though this process itself does not explicitly show CP 
violation.

\par
We show that the neutrino mixing angles are constrained from the presence of 
CP violation.  Here we follow the method given in \cite{nishiura}.
\par
From Eq.(3) it follows that
\begin{equation}
\langle m_{\nu} \rangle^2=(m_1c_1^2c_3^2-m_2s_1^2c_3^2cos2\beta 
'-m_3s_3^2cos2\rho ')^2
+(m_2s_1^2c_3^2sin2\beta '+m_3s_3^2sin2\rho ')^2
\end{equation}    
Rewriting $cos2\rho '$ and $sin2\rho '$ by $tan\rho '$, we can 
consider
Eq.(5) 
as an equation of $tan\rho '$,
\begin{equation}
a_{+\beta}tan^2\rho '+b_{\beta}tan\rho '+a_{-\beta}=0.
\end{equation}  
Here $a_{\pm \beta}$ and $b_{\beta}$ are defined by
\begin{eqnarray}
a_{\pm \beta}&\equiv& 4sin^2\beta 'm_2s_1^2c_3^2(m_1c_1^2c_3^2\pm 
m_3s_3^2)+(m_1c_1^2c_3^2-m_2s_1^2c_3^2\pm m_3s_3^2)^2-\langle 
m_{\nu} 
\rangle^2\\
b_\beta&\equiv& 4m_2m_3s_1^2s_3^2c_3^2sin2\beta '.\nonumber
\end{eqnarray}
So, the discriminant $D$ for 
Eq.(6)
must 
satisfy the following inequality:
\begin{eqnarray}
D&\equiv& b_\beta^2-4a_{+\beta}a_{-\beta}\nonumber\\
&=&4^3(m_1c_1^2c_3^2)^2(m_2s_1^2c_3^2)^2(f_+-sin^2\beta 
')(sin^2\beta
'-f_-)\geq 
0,
\end{eqnarray}
where 
\begin{equation}
f_{\pm}\equiv {(\langle m_{\nu} \rangle\pm 
m_3s_3^2)^2-(m_1c_1^2c_3^2-m_2s_1^2c_3^2)^2\over
4m_1m_2c_1^2s_1^2c_3^4}.
\end{equation}
So we obtain
\begin{equation}
f_-\leq sin^2\beta '\leq f_+.
\end{equation}
It follows from Eq.(10) that
\begin{equation}
f_-\leq 1,~~ f_+\geq 0.
\end{equation}
Quite analogously, rewriting $cos2\beta '$ and $sin2\beta '$ by 
$tan\beta'$, and considering Eq.(5) as an equation of $tan\beta '$, we 
obtain other inequalities,
\begin{equation}
g_-\leq sin^2\rho '\leq g_+. 
\end{equation}
Here
\begin{equation}
g_{\pm}\equiv {(\langle m_{\nu} \rangle\pm 
m_2s_1^2c_3^2)^2-(m_1c_1^2c_3^2-m_3s_3^2)^2\over
4m_1m_3c_1^2s_3^2c_3^2}.
\end{equation}
So we get 
\begin{equation}
g_-\leq 1,~~ g_+\geq 0.
\end{equation}
The conditions (11) and (14) are consistency conditions. CP violating
area is given by the more stringent condition
\begin{equation}
0\leq f_-\leq sin^2\beta '\leq f_+\leq 1 \quad
or \quad
0\leq g_-\leq sin^2\rho '\leq g_+\leq 1.
\end{equation}
From the inequalities (11) and (14), we obtain the allowed region of the mixing 
angles in the $s_1^2$ versus $s_3^2$ plane once the neutrino masses $m_i$ and the 
"averaged " neutrino  mass $\langle m_{\nu}\rangle $ are known.  The magnitude 
of $\langle m_{\nu}\rangle $ is experimentally unknown at present. 
The neutrino masses may be safely ordered as \(m_1 \le m_2 \le m_3\).
So in the following discussions we consider the three cases:

a) $\langle m_\nu \rangle \leq m_1$,
\par

b) $m_1 \leq \langle m_\nu \rangle \leq m_2$,
\par
and
\par
c) $m_2 \leq \langle m_\nu \rangle \leq m_3$.

Note that the definition of $\langle m_{\nu}\rangle $ in Eq.(1) and the Schwartz 
inequality leads us to
\begin{equation}
\langle m_\nu \rangle \leq \sum _{j=1}^{3} m_j|U_{ej}^2|\leq m_3\sum
_{j=1}^{3} 
|U_{ej}^2|=m_3,
\end{equation}
so $\langle m_\nu \rangle$ can not be larger than $m_3$.  The allowed regions in the 
$s_1^2$ versus $s_3^2$ plane for each case (a), (b) and (c) are obtained from 
Eqs.(11) and (14), and are shown in Fig.2.
\begin{center}
Fig.2
\end{center}
From Fig.2, we obtain the upper bound on $s_3^2$ as
\begin{equation}
s_3^2\leq {m_2+\langle m_{\nu} \rangle \over m_3+m_2}
\end{equation}
for any case.  The CP violating areas in the $s_1^2$ versus $s_3^2$ plane given 
by Eq.(15) are also indicated by the oblique lines in Fig.2 for each case 
(a), (b) and (c).  The above case (a) was considered also in \cite{nishiura} 
and\cite{minakata}.  In \cite{nishiura}, the 
representation for the mixing matrix adopted by 
Cabibbo-Kobayashi-Maskawa was used. In \cite{minakata}, only 
the limiting case where all the neutrino masses are degenerate ($m_1=m_2=m_3$) 
was discussed.  It should be noted that we consider the cases (b) and (c) in 
addition to (a) and 
that no condition on the neutrino masses has been imposed so far.

The above mentioned method is not restricted to Majorana particles but is widely 
applicable.  Next, we consider the constraint from the neutrino oscillation 
experiment at CHORUS\cite{chorus} and see how this method also gives the allowed region in $s_1^2$ 
versus $s_3^2$ plane.  We assume here $\delta m_{31}^2\equiv m_3^2-m_1^2\sim 
\delta  m_{21}^2\equiv m_2^2-m_1^2\gg\delta m_{32}^2\equiv m_3^2-m_2^2.$

In this case the approximate oscillation probability is given
by\cite{tanimoto}
\begin{equation}
P(\nu_{\mu}\rightarrow\nu_{\tau})=4|U_{\mu 1}|^2|U_{\tau 1}|^2sin^2
({\delta m_{31}^2L\over 4E_\nu}).
\end{equation}
Substituting the expression of Eq.(2) into Eq.(18), we obtain the 
following equation w.r.t. $cos\phi$,
\begin{eqnarray}
 A & \equiv & \frac{P(\nu_\mu\rightarrow\nu_\tau)}
                {4\sin^2(\frac{\delta m_{31}^2}{4E_\nu}L)}\nonumber\\
                  & = & a_+ \cos^2\phi-2b\cos\phi+a_-\\
                  & \equiv & f(\cos\phi).\nonumber
\end{eqnarray}

Here
\begin{eqnarray}
 a_+&\equiv &-(2s_1s_2s_3c_1c_2)^2,\nonumber\\
 a_-&\equiv 
&(s_1^2c_2^2+c_1^2s_2^2s_3^2)(s_1^2s_2^2+c_1^2c_2^2s_3^2),\\
 b&\equiv 
 &s_1s_2s_3c_1c_2(s_1^2-c_1^2s_3^2)(c_2^2-s_2^2).\nonumber
\end{eqnarray}
The oscillation process does not distinguish Majorana neutrino from Dirac one, 
and only $\phi$ phase takes place.  Firstly the discriminant D of Eq(19) leads 
to\begin{eqnarray}
   D & \equiv & b^2-a_+(a_--A)\nonumber\\
       & = & (s_1s_2s_3c_1c_2)^2
             [(s_1^2+c_1^2s_3^2)^2-4A] \ge 0.
\end{eqnarray}
Thus we obtain
\begin{equation}
  \frac{(s_1^2+c_1^2s_3^2)^2}{4} \ge A,\label{m1}
\end{equation}
which is irrelevant to \(\theta_{23}\).
Therefore the CHORUS data on $P(\nu_\mu \to \nu_\tau)$ ($\equiv P_{CHORUS}$) constrain 
the allowed region in the $s_1^2$ versus $s_3^2$ plane (Fig.3).
\begin{center}
Fig.3
\end{center}
Unfortunately, we have only the upper bound on the $P_{CHORUS}$, $P_{CHORUS} < 2.5 \times 10^{-3}$ \cite{Shibuya}. Setting \(L=600m\) (the midpoint of the maximum length, 800m and the minimum length, 400m), \(E=27GeV\) and 
$\delta m_{31}^2 \sim \delta m_{21}^2 = 6 eV^2 \gg 
\delta m_{32}^2$, we have \(A<0.022\). The broken line is the trajectory of
\begin{equation}
\frac{(s_1^2+c_1^2s_3^2)^2}{4} =0.022,
\end{equation}
and the allowed region is the upper part from the broken line.  If the $P_{CHORUS}$ 
gives the lower value, the broken line moves downward to extend the allowed region.  The shaded areas in 
Fig.3 are those of Fig.2(b) under the assumption that $\delta m_{31}^2 \sim \delta 
m_{21}^2 = 6 eV^2 \gg \delta m_{32}^2\ $ and \(m_1 \ll \langle m_\nu \rangle\) 
with possible $\langle m_{\nu} \rangle$ 
values. 
The more stringent constraints, though they depend on 
\(\theta_{23}\), are also obtained from Eq.(19). \(a_+\) is negative definite 
and 
\(f(\pm1)\)
are positive definite. Therefore from the condition 
that $-1\leq cos\phi \leq 1$ we 
obtain the following inequalities:
\begin{description}
\item[Case a-1 : ] 
\(
 0 \le \frac{(s_2^2-c_2^2)(s_1^2-c_1^2s_3^2)}{4s_1s_2s_3c_1c_2} \le 1
\)
\begin{equation}
(s_1c_2-c_1s_2s_3)^2(s_1s_2+c_1c_2s_3)^2 \le 
\frac{P(\nu_\mu\rightarrow\nu_\tau)}{4\sin^2(\frac{\delta
m_{31}^2}{4E_\nu}L)} 
      \le \frac{1}{4}(s_1^2+c_1^2s_3^2)^2 \label{a1}
\end{equation}
\item[Case a-2 : ]
\(
1<\frac{(s_2^2-c_2^2)(s_1^2-c_1^2s_3^2)}{4s_1s_2s_3c_1c_2}
\)
\begin{equation}
(s_1c_2-c_1s_2s_3)^2(s_1s_2+c_1c_2s_3)^2\le
\frac{P(\nu_\mu\rightarrow\nu_\tau)}{4\sin^2(\frac{\delta
m_{31}^2}{4E_\nu}L)}
\le (s_1c_2+c_1s_2s_3)^2(s_1s_2-c_1c_2s_3)^2 \label{a2}
\end{equation}
\item[Case b-1 : ]
\(
-1 \le \frac{(s_2^2-c_2^2)(s_1^2-c_1^2s_3^2)}{4s_1s_2s_3c_1c_2} \le 0
\)
\begin{equation}
(s_1c_2+c_1s_2s_3)^2(s_1s_2-c_1c_2s_3)^2 \le 
\frac{P(\nu_\mu\rightarrow\nu_\tau)}{4\sin^2(\frac{\delta
m_{31}^2}{4E_\nu}L)} 
      \le \frac{1}{4}(s_1^2+c_1^2s_3^2)^2 \label{b1}
\end{equation}
\item[Case b-2 : ]
\(
\frac{(s_2^2-c_2^2)(s_1^2-c_1^2s_3^2)}{4s_1s_2s_3c_1c_2}<-1
\)
\begin{equation}
(s_1c_2+c_1s_2s_3)^2(s_1s_2-c_1c_2s_3)^2\le
\frac{P(\nu_\mu\rightarrow\nu_\tau)}{4\sin^2(\frac{\delta
m_{31}^2}{4E_\nu}L)}
\le (s_1c_2-c_1s_2s_3)^2(s_1s_2+c_1c_2s_3)^2 \label{b2}
\end{equation}
\end{description}

As we have mentioned, we have experimentally only the upper bound of 
$P(\nu_{\mu}\rightarrow \nu_{\tau})$ at present.  So the more meaningful inequalities than Eq.(22)
comes from the lower bounds of Eqs.(24) \(\sim\) (27).
Namely we have\\
\\
Case a
\begin{equation} 
(s_1c_2-c_1s_2s_3)^2(s_1s_2+c_1c_2s_3)^2\leq\frac{P(\nu_\mu\rightarrow\nu_\tau)}
{4\sin^2(\frac{\delta
m_{31}^2}{4E_\nu}L)}
 \quad for\quad (s_2^2-c_2^2)(s_1^2-c_1^2s_3^2) \ge 0
\end{equation}
Case b
\begin{equation}
(s_1c_2+c_1s_2s_3)^2(s_1s_2-c_1c_2s_3)^2\leq\frac{P(\nu_\mu\rightarrow\nu_\tau)}
{4\sin^2(\frac{\delta
m_{31}^2}{4E_\nu}L)} \quad for \quad (s_2^2-c_2^2)(s_1^2-c_1^2s_3^2) \le 0
\end{equation}
From these inequalities (28) and (29), we obtain another allowed region in the 
$s_1^2$ versus $s_3^2$ plane for a fixed value of $\theta_2$.  Using $\delta 
m_{31}^2 \sim \delta m_{21}^2 = 6 eV^2 \gg \delta m_{32}^2$ and 
$A < 0.022$, we show the allowed regions for 
$\theta_2=0, {\pi\over 24}, {\pi\over 12},\cdots ,{\pi\over 2}$ in Fig.4.
\begin{center}
Fig.4
\end{center}
Lastly we comment on another neutrino less process of 
\(\mu^-\)-\(e^+\) conversion [11]. (Fig.5)
\begin{center}
Fig.5
\end{center}
In this case the averaged neutrino mass which will be determined experimentally 
is given by

\begin{equation}
	\langle m_\nu \rangle_{\mu^-e^+} \equiv 
	\left|\sum_{i=1}^{3}{m_i U_{ei}^* U_{\mu i}^*}\right|=
	\left|\sum_{i=1}^{3}{m_i U_{ei} U_{\mu i}}\right|.
\end{equation}
Substituting the expression of Eq.(2) into Eq.(30), we obtain
\begin{eqnarray}
\langle m_\nu \rangle_{\mu^-e^+}^{\ 2} 
	&=& m_{1}^{2} (c_{1}^{2} c_{2}^{2} c_{3}^{2}
s_{1}^{2}+2 \cos\phi c_{1}^{3} {c_2} c_{3}^{2} {s_1} {s_2}
{s_3}+c_{1}^{4} c_{3}^{2} s_{2}^{2} s_{3}^{2}) \nonumber \\
&&+m_{2}^{2} (c_{1}^{2} c_{2}^{2} c_{3}^{2} s_{1}^{2}-2 \cos\phi  {c_1}
{c_2} c_{3}^{2} s_{1}^{3} {s_2} {s_3}+c_{3}^{2} s_{1}^{4} s_{2}^{2}
s_{3}^{2}) \nonumber \\
&&+m_{3}^{2} c_{3}^{2} s_{2}^{2} s_{3}^{2} \nonumber \\
&&+{m_1} {m_2} (-2 \cos (2 \beta ) c_{1}^{2} c_{2}^{2} c_{3}^{2}
s_{1}^{2}-2 \cos (2 \beta -\phi ) c_{1}^{3} {c_2} c_{3}^{2} {s_1} {s_2}
{s_3} \nonumber \\
&&\quad +2 \cos (2 \beta +\phi ) {c_1} {c_2} c_{3}^{2} s_{1}^{3} {s_2}
{s_3}+2 \cos (2 \beta ) c_{1}^{2} c_{3}^{2} s_{1}^{2} s_{2}^{2}
s_{3}^{2}) \nonumber \\
&&+{m_1} {m_3} (-2 \cos (2 \rho -\phi ) {c_1} {c_2} c_{3}^{2} {s_1}
{s_2} {s_3}-2 \cos (2 \rho -2 \phi ) c_{1}^{2} c_{3}^{2} s_{2}^{2}
s_{3}^{2}) \nonumber \\
&&+{m_2} {m_3} (2 \cos (2 \beta -2 \rho +\phi ) {c_1} {c_2} c_{3}^{2}
{s_1} {s_2} {s_3} \nonumber \\
&&\quad - 2 \cos (2 \beta -2 \rho +2 \phi ) c_{3}^{2} s_{1}^{2}
s_{2}^{2} s_{3}^{2}).
\end{eqnarray}
In contrast to the neutrinoless double beta decay, 
all the mixing angles and the phase parameters appear. 
So if we assume one of the phases and $\theta_{23}$, we 
can develop the same argument as that in the neutrinoless double beta decay. 

In conclusion, we have proposed the new method to constrain the neutrino mixing angles from the observed data of \(\langle m_\nu \rangle\). Our method, however, is widely valid and have been applied to the neutrino oscillation, having given the new constraints from the observed data of \(P(\nu_\mu \to \nu_\tau)\). 
Our method will be applied to the other decay and oscillation processes.

\ \\
Acknowledgement:

We are grateful to H. Shibuya for informing us the most recent result of 
CHORUS group.
\clearpage

\vfill\eject





\clearpage
Figure Captions
\par
\begin{description}
   \item[Fig.1:]

	Feynman diagram of  neutrinoless double beta decay.

   \item[Fig.2:]

	The allowed region on $sin^2\theta_{12}$ versus 
	$\sin^2\theta_{31}$
	plane by the neutrinoless double beta decay is 
	given by the shaded areas in the respective case: 
	\begin{center}
	(a) $\langle m_\nu \rangle \leq m_1$ \quad
	(b) $m_1 \leq \langle m_\nu\rangle \leq m_2$ \quad 
	(c) $m_2 \leq \langle m_\nu \rangle \leq m_3$
	\end{center}

	In the allowed region, CP-violating area is specially indicated by 
	the oblique lines.

   \item[Fig.3:]

	Each allowed region by the neutrinoless double beta decay 
	in the respective case:
	\begin{center}
	\(\langle m_\nu \rangle = 0.4,\, 0.8,\, 1.2,\, 1.6,\, 2.0,\, 2.4eV\)
	\end{center}
	under the assumption that 
	\(\delta m_{31}^2 \sim \delta m_{21}^2 = 6 eV^2 \gg \delta m_{32}^2\) 
	and \(m_1\ll\langle m_\nu\rangle\).
	Broken line is given by Eq.(23). 
	The upper part from the broken line is allowed by 
	the neutrino oscillation experiment at CHORUS.

   \item[Fig.4:]

	The allowed regions by 
	the inequalities of Eqs.(28) and (29) under the condition that
	\(P_{CHORUS}<2.5\times10^{-3}\), 
	\(\delta m_{31}^2 \sim \delta m_{21}^2 = 6 eV^2 \gg \delta m_{32}^2\) 
	with given \(\theta_2\).
	\begin{center}
	\parbox{11cm}{
	  (a)\(\theta_2=0, \frac{\pi}{2}\) \quad 
	  (b)\(\theta_2=\frac{\pi}{24}, \frac{11\pi}{24}\) \quad 
	  (c)\(\theta_2=\frac{\pi}{12}, \frac{5\pi}{12}\) \quad 
	  (d)\(\theta_2=\frac{\pi}{8}, \frac{3\pi}{8}\)\\
	  (e)\(\theta_2=\frac{\pi}{6}, \frac{\pi}{3}\) \quad 
	  (f)\(\theta_2=\frac{5\pi}{24}, \frac{7\pi}{24}\) \quad 
	  (g)\(\theta_2=\frac{\pi}{4}\)}
	\end{center}

    \item[Fig.5:]

	Feynman diagram of the \(\mu^--e^+\) conversion.

\end{description}

\clearpage
\begin{figure}[htbp]
	\begin{center}
	\leavevmode
	\epsfile{file=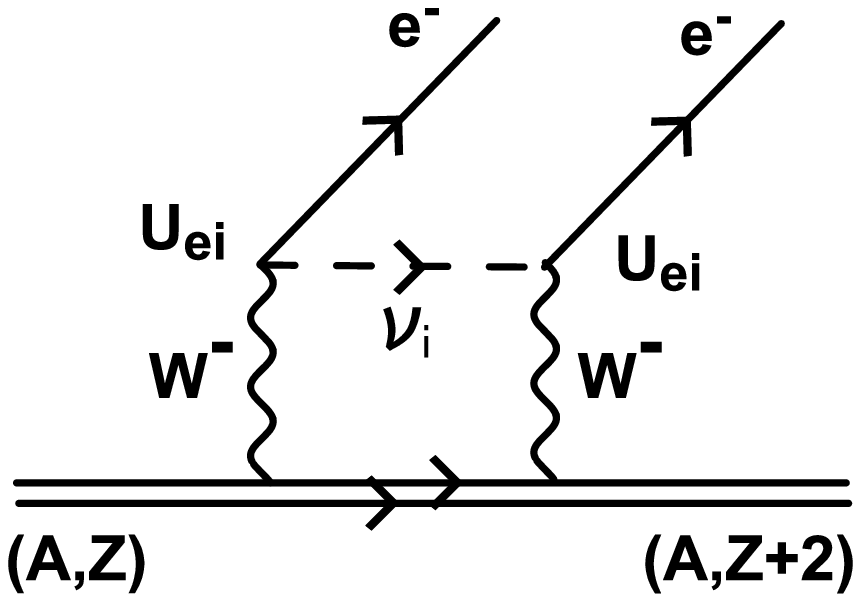,width=13.12cm,height=9.6536cm}\\
\ \\
	{\Huge Fig.1}
	\end{center}
\end{figure}

\clearpage
\begin{figure}[hbtp]
	\begin{center}
	\leavevmode
	\epsfile{file=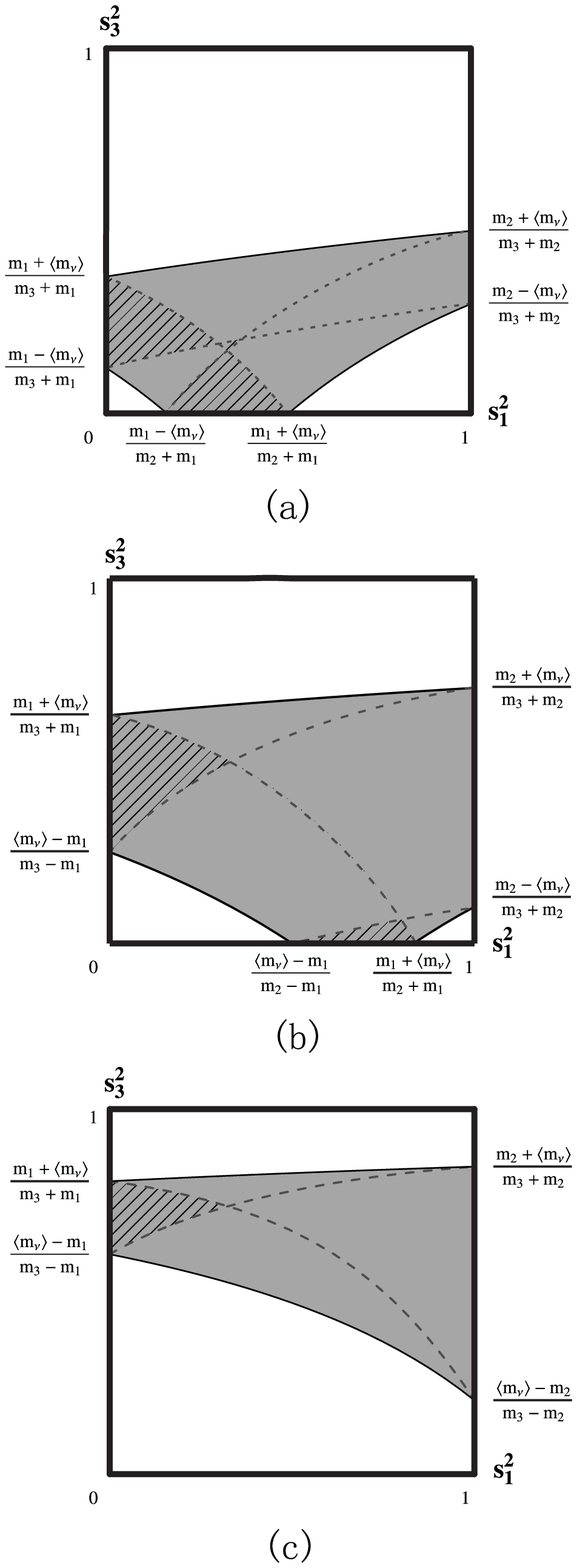,width=7.2cm,height=19.84cm}\\
\ \\
	{\Huge Fig.2}
	\end{center}
\end{figure}

\clearpage
\begin{figure}[htbp]
	\begin{center}
	\leavevmode
	\epsfile{file=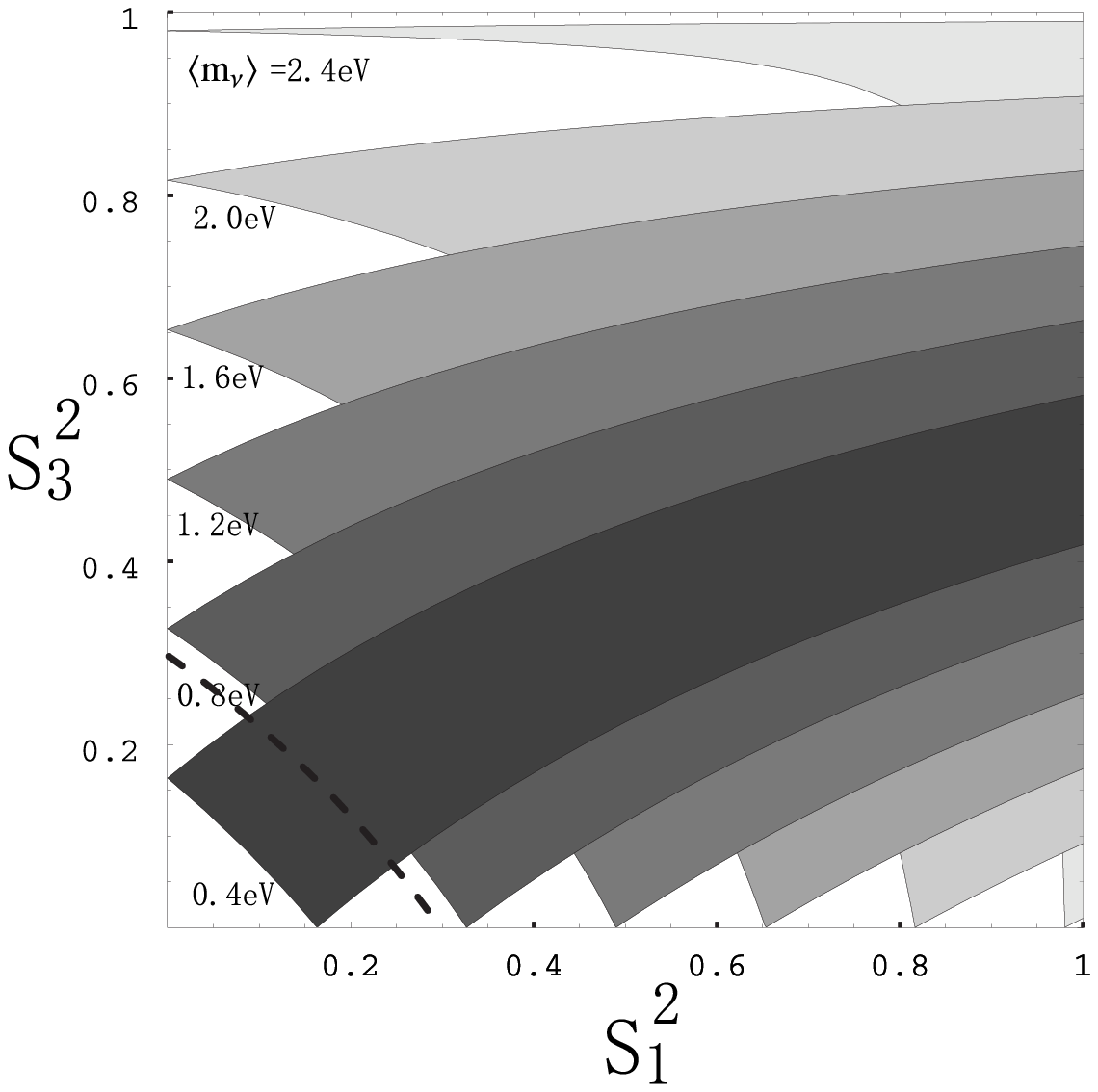,width=12.225cm,height=12.3cm}\\
\ \\
	{\Huge Fig.3}
	\end{center}
\end{figure}

\clearpage
\begin{figure}[htb]
	\begin{center}
	\leavevmode
	\epsfile{file=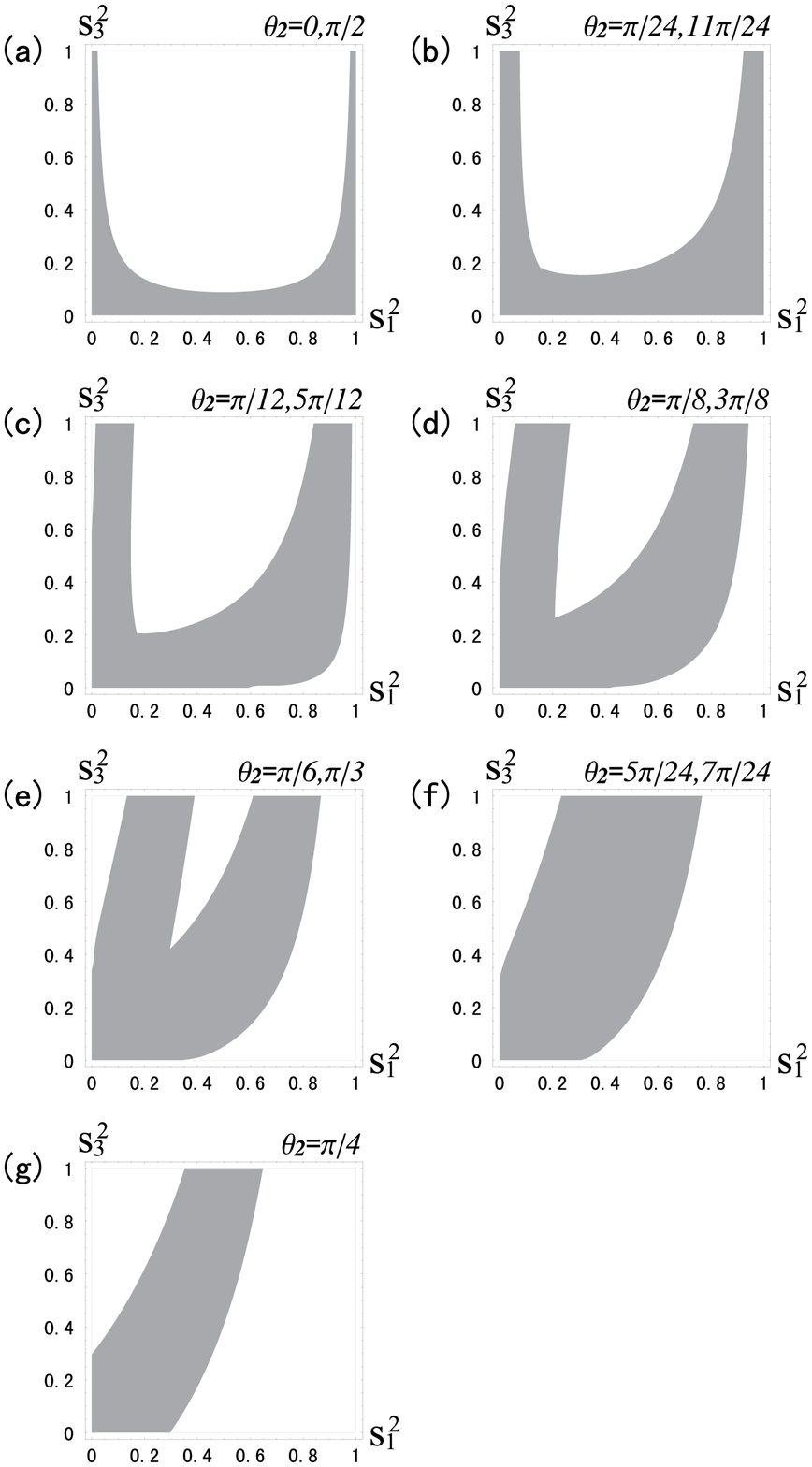,width=11.013cm,height=19.973cm}\\
\ \\
	{\Huge Fig.4}
	\end{center}
\end{figure}

\clearpage
\begin{figure}[htb]
	\begin{center}
	\leavevmode
	\epsfile{file=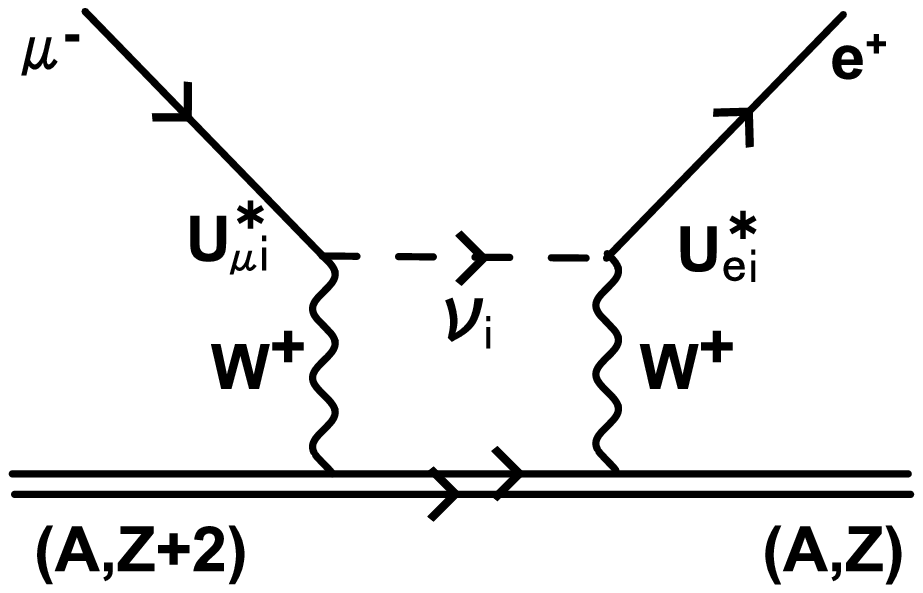,width=13.12cm,height=9.6536cm}\\
\ \\
	{\Huge Fig.5}
	\end{center}
\end{figure}


\begin{thebibliography} {99}
\bibitem{Kamioka}
K.S. Hirata et. al., Phys. Rev Lett. {\bf 63} 6 (1989); N.
Hata,S.Bludman
and
P. Langacker, Phys. Rev.{\bf D49} 3622 (1994); N. Hata and P.Langacker,
Phys.Rev.{\bf D50} 632 (1994)
\bibitem{Homestake}
R. Davis Jr., Proc. of the 6'th International Workshop on Neutrino
Telescope, Venezia, March 1994, edited by M. Baldo Ceolin.
\bibitem{Gallex}
P. Anselmann et.al., Phys.Lett {\bf B285} 376 (1992), ibid {\bf
B357} 237 (1995).
\bibitem{Sage}
A.I. Abazov et. al., Phys. Rev. Lett. {\bf 67} 3332 (1991).;J.N.
Abdurashitov et. al., Phys. Lett. {\bf B328} 234 (1994).
\bibitem{doi}
M. Doi, T. Kotani, H. Nishiura, K. Okuda and E. Takasugi, Phys. Lett. {\bf 102B} 
323 (1981); M. Doi, T. Kotani and E. Takasugi, Prog. Theor. Phys. Suppl. {\bf 
83} 1 (1985); W.C. Haxton and G.J. Stophenson Jr., Prog. Part. Nucl. Phys. {\bf 
12} 409 (1984).
\bibitem{nishiura}
H. Nishiura, Phys. Lett. {\bf 157B} 442 (1985).
\bibitem{minakata}
H. Minakata, hep-ph/9612259.
\bibitem{chorus}
K. Winter, Nucl. Phys. {\bf B38} (Proc. Suppl.) 211 (1995).
\bibitem{tanimoto}
M. Tanimoto, Ehime University Preprint EHU-96-2

\bibitem{Shibuya}
Private communication with H. Shibuya(July 1997). See also the Preprint,
"The CHORUS Neutrino Oscillation Search Experiment" CERN-PPE/96-196.

\bibitem{Kamal}
A. N. Kamal and J. N. Ng, Phys. Rev. {\bf D20} 2269 (1979);
J. D. Vergados and M. Ericson, Nucl. Phys. {\bf B195} 262 (1982);
G. K. Leontaris and J. D. Vergados, Nucl. Phys. {\bf B224} 137 (1983);
M. Doi,T. Kotani and E. Takasugi, Prog. Theor. Phys. Suppl. {\bf 83} 1 (1985).

\end{thebibliography}
\end{document}